\def\year{2020}\relax
\documentclass[letterpaper]{article} 
\usepackage{aaai20}  
\usepackage{times}  
\usepackage{helvet} 
\usepackage{courier}  
\usepackage[hyphens]{url}  
\usepackage{graphicx} 
\usepackage{graphicx}  
\usepackage[abs]{overpic}
\usepackage{amsmath}
\usepackage{color}
\title{An Integrated Enhancement Solution for 24-hour Colorful Imaging}
\author{
	Feifan Lv,\textsuperscript{\rm 1}
	Yinqiang Zheng,\textsuperscript{\rm 2}
	Yicheng Li,\textsuperscript{\rm 1}
	Feng Lu\textsuperscript{\rm 1,3,}\thanks{Corresponding Author. \newline \indent~ This work is partially supported by the National Natural Science Foundation of China (NSFC) under Grant 61972012 and Grant 61732016.}\\
	\textsuperscript{\rm 1}State Key Laboratory of VR Technology and Systems, School of CSE, Beihang University, Beijing, China \\
	\textsuperscript{\rm 2}National Institute of Informatics, Tokyo, Japan  \hspace{28pt}
	\textsuperscript{\rm 3}Peng Cheng Laboratory, Shenzhen, China \\
	lvfeifan@buaa.edu.cn, yqzheng@nii.ac.jp, liyicheng@buaa.edu.cn, lufeng@buaa.edu.cn
}

\begin{document}
	
	\maketitle
	\begin{abstract}
		The current industry practice for $24$-hour outdoor imaging is to use a silicon camera supplemented with near-infrared (NIR) illumination. This will result in color images with poor contrast at daytime and absence of chrominance at nighttime. For this dilemma, all existing solutions try to capture RGB and NIR images separately. However, they need additional hardware support and suffer from various drawbacks, including short service life, high price, specific usage scenario, etc. In this paper, we propose a novel and integrated enhancement solution that produces clear color images, whether at abundant sunlight daytime or extremely low-light nighttime. Our key idea is to separate the VIS and NIR information from mixed signals, and enhance the VIS signal adaptively with the NIR signal as assistance. To this end, we build an optical system to collect a new VIS-NIR-MIX dataset and present a physically meaningful image processing algorithm based on CNN. Extensive experiments show outstanding results, which demonstrate the effectiveness of our solution.
	\end{abstract}
	
	\section{Introduction}
	Many practical imaging systems, such as surveillance systems, require continuous and stable operation. High-quality images are expected to be captured in $24$ hours. As a result, ambient illumination change becomes a non-negligible factor. Common cameras usually work well under abundant daylight, but perform badly at low-light nighttime. Simply using a flash unit or increasing the exposure time are impractical as they will change the tone of imaging, exposure coverage scope and cause image blur.
	
	\begin{figure}[t]
		\begin{center}
			\includegraphics[scale=0.62]{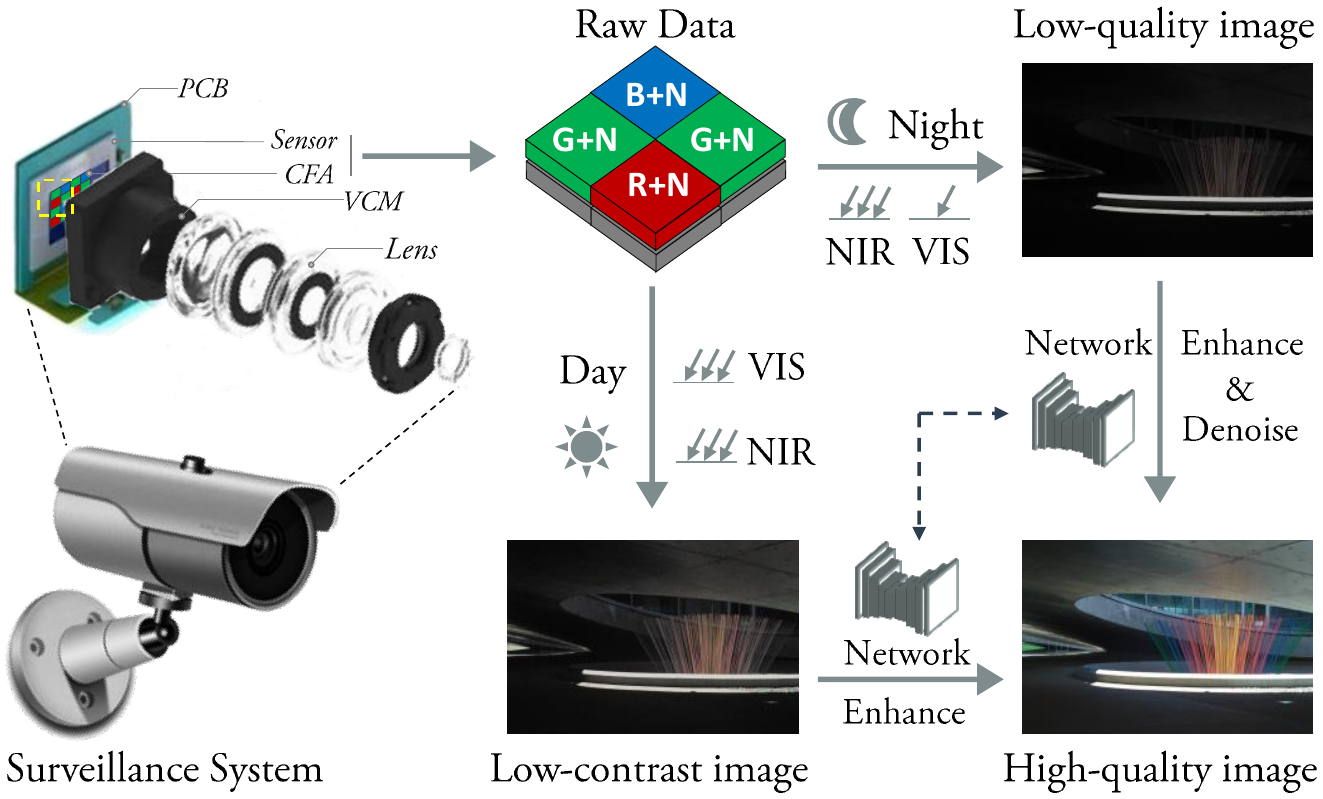}
		\end{center}
		\caption{The proposed 24-hour imaging solution using only a Silicon sensor. VIS and NIR signals are automatically separated and processed to produce constantly clear and colorful images during both daytime and nighttime.}
		\label{fig_diagram}
	\end{figure}
	
	To overcome the limitation, the majority of existing solutions use additional near-infrared (NIR) illumination to light up the object and capture it. However, NIR is a double-edged sword for those systems. On one hand, it utilizes the sensitivity of Silicon sensors around 700nm-950nm, allowing visual information acquisition in the dark. On the other hand, it affects the capture of visible spectrum (VIS) information and results in degraded color images.
	
	For this dilemma, advanced solutions tend to capture NIR and VIS signals separately. Three typical hardware-assisted solutions include using steerable IR-Cut filters, using specific color filter arrays (CFAs) and using two different imaging sensors. In the first case, an IR-Cut filter is automatically or periodically equipped under adequate illumination and is removed otherwise. However, this requires additional illumination intensity sensor and also reduces the service life.
	In the second case, specific CFAs directly capture NIR and VIS information, but the image resolution and quality are highly limited. For the last option, involving two cameras raises the cost and also faces difficulty in meticulous calibration. Overall, all existing solutions that try to capture both NIR and VIS signals still show limitations.
	
	Different from existing solutions, in this paper we propose to separate VIS and NIR signals directly from the mixed signals captured by a single sensor without any additional hardware, as illustrated in Figure~\ref{fig_diagram}. The required hardware is only a common Silicon sensor without IR-Cut filters. With our proposed algorithm, our solution not only significantly reduces the cost, but also produces clear and colorful images at both daytime and nighttime.
	
\begin{figure}[t]
		\begin{center}
			\includegraphics[scale=0.65]{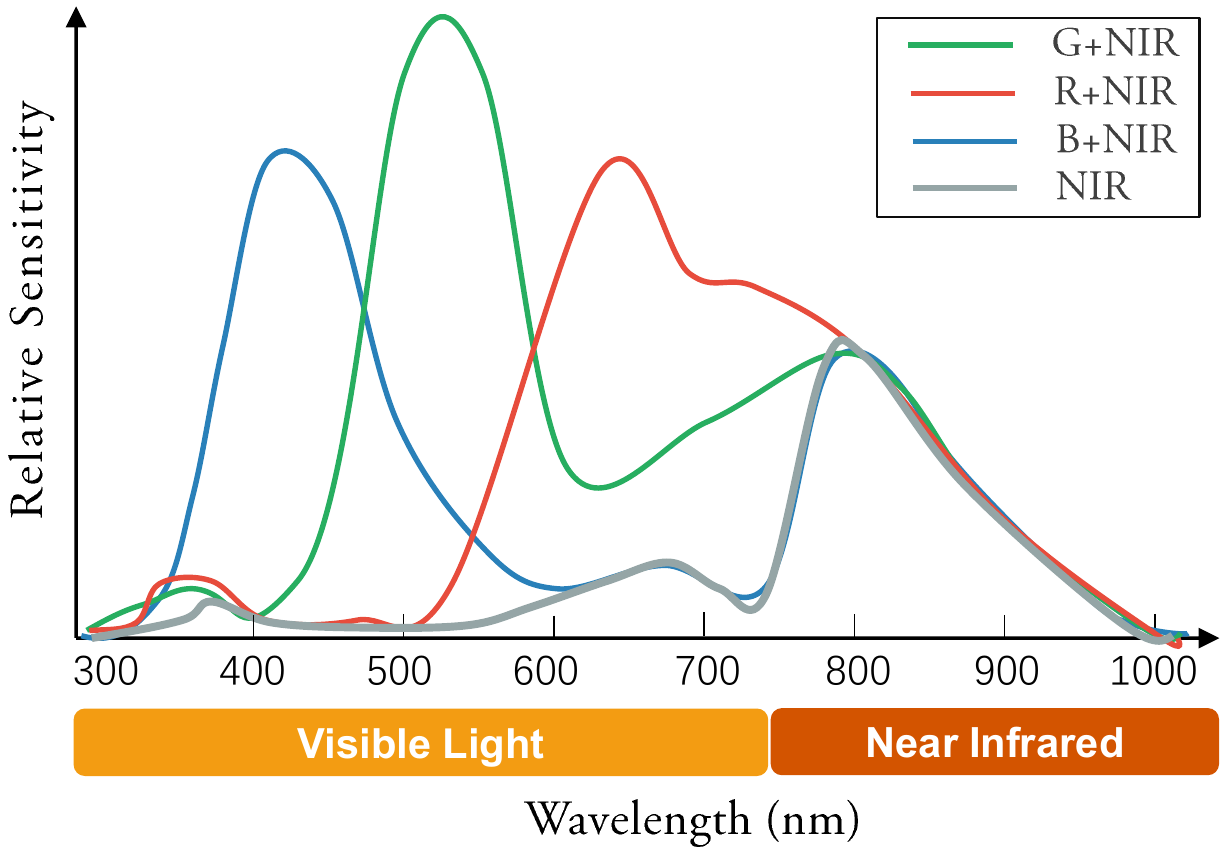}
		\end{center}
		\caption{The relative spectral sensitivity after the CFA of our imaging system without hot-mirror. The gray curve represents the sensitivity of the latent pure NIR signal. The other three colored curves are NIR and VIS mixed signals.}
		\label{fig_curve}
	\end{figure}

	Our key idea lies in the proper use of the wide range, \textit{i.e.}, $300$-$950$nm, sensitivity of a Silicon senor, to capture the adequate information of VIS and NIR components, as shown in Figure~\ref{fig_curve}.
	Consequently, we propose to separate the VIS and NIR signal automatically from the mixed signal, and use the NIR signal as guidance for VIS image enhancement and restoration. To this purpose, we design an optical imaging system and build a novel VIS-NIR-MIX image dataset, allowing the learning of a physically meaningful RAW-to-VIS mapping network.
	
	Both qualitative and quantitative experiments have shown that our proposed solution is capable of generating colorful and clear images in both abundant sunlight and extremely low-light scenes, which outperforms the existing solutions.
	Overall, our main contributions are threefold:
	\begin{itemize}
		\item We propose an integrated enhancement solution for $24$-hour high-quality imaging by using only a common Silicon sensor without additional hardware.
		\item We design a prototype imaging system and build a new dataset with aligned RAW/VIS/NIR images, allowing RAW-to-VIS mapping in a more accurate way.
		\item We propose and implement our image processing model by an end-to-end network. Promising results on high-quality image generation are provided.
	\end{itemize}

	\section{Related work}
	Imaging under unstable illumination is the major challenge for many imaging systems. Most existing researches focus on specific cases, rather than providing an integrated $24$-hour colorful imaging solution. According to the used information, they can be roughly divided into three categories: using only VIS signals, using only NIR signals and using both VIS/NIR signals.
	In this section, we will briefly overview the most relevant works.
	
	{\bf Using only VIS signals.} Enhancement and restoration of VIS images has been studied intensively. Existing enhancement methods mainly contain histogram equalization (HE) based methods~\cite{nakai2013dheci}, Retinex theory based methods~\cite{guo2017lime} and learning based methods~\cite{chen2018learning,lv2018mbllen,lv2019AgLLNet}. Existing denoising methods are massive, typically including filter based methods~\cite{dabov2007color}, effective prior based methods~\cite{xu2018external}, low rank based methods~\cite{yair2018multi}, sparse coding based methods~\cite{xu2018trilateral} and learning based methods~\cite{remez2017deep}.
	Although these researches have greatly promoted the image enhancement and denoising techniques, it's still far from the requirements of 24-hour imaging systems.
	
	{\bf Using only NIR signals.} Some researches~\cite{limmer2016infrared,suarez2017infrared} consider directly learning the NIR to RGB mapping to avoid the effects of unstable ambient VIS illumination. However, NIR images and VIS images are not exactly the same, which will result in unrealistic results. More fundamentally, the color of object is often inherently ambiguous, which means that it is difficult to accurately estimate the color information from the NIR signal without additional constraints. Therefore, such colorization methods are usually used for specific purposes rather than high-quality imaging systems.
	
	{\bf Using both VIS and NIR signals.} Since NIR images have various advantages, \textit{e.g.}, haze free, solutions using both VIS and NIR signals have become the industry standard for $24$-hour imaging systems. There are three common types: using steerable hot-mirrors, using specific CFAs and using two cameras.
	In the first scenario, switching hot-mirrors ensures colorful imaging at daytime while almost gray images obtained at nighttime. Besides the defect of chromaticity loss, automatic switching will also raise the cost and reduce service life.
	For the second case, the special CFA contains NIR components, which makes it can capture RGB and NIR signals simultaneously. Many researches focus on the RGB/NIR separation~\cite{teranaka2016single}, RGB/NIR jointly restoration~\cite{park2016color} and low-light imaging~\cite{yamashita2017low} by using special CFAs. This special CFA trades the low-resolution NIR signal at the expense of color image resolution, which will affect the image quality. Besides, the advantages of these special CFAs will translate into disadvantages compared with normal bayer filters at daytime.
	As for the last strategy, it can be seen as an upgraded version of the second case, since two cameras can simultaneously capture high-resolution NIR and VIS images. This will provide prerequisites for researches on low-light RGB/NIR processing~\cite{sugimura2015enhancing} and RGB/NIR jointly restoration~\cite{yan2013cross}. Nevertheless, two cameras will raise the cost and face the serious problem of meticulous calibration.
	
	\section{Imaging in Both Daytime and Nighttime: Analysis and an Integrated Solution}
	
\begin{figure}[t]
		\begin{center}
			\includegraphics[scale=0.5]{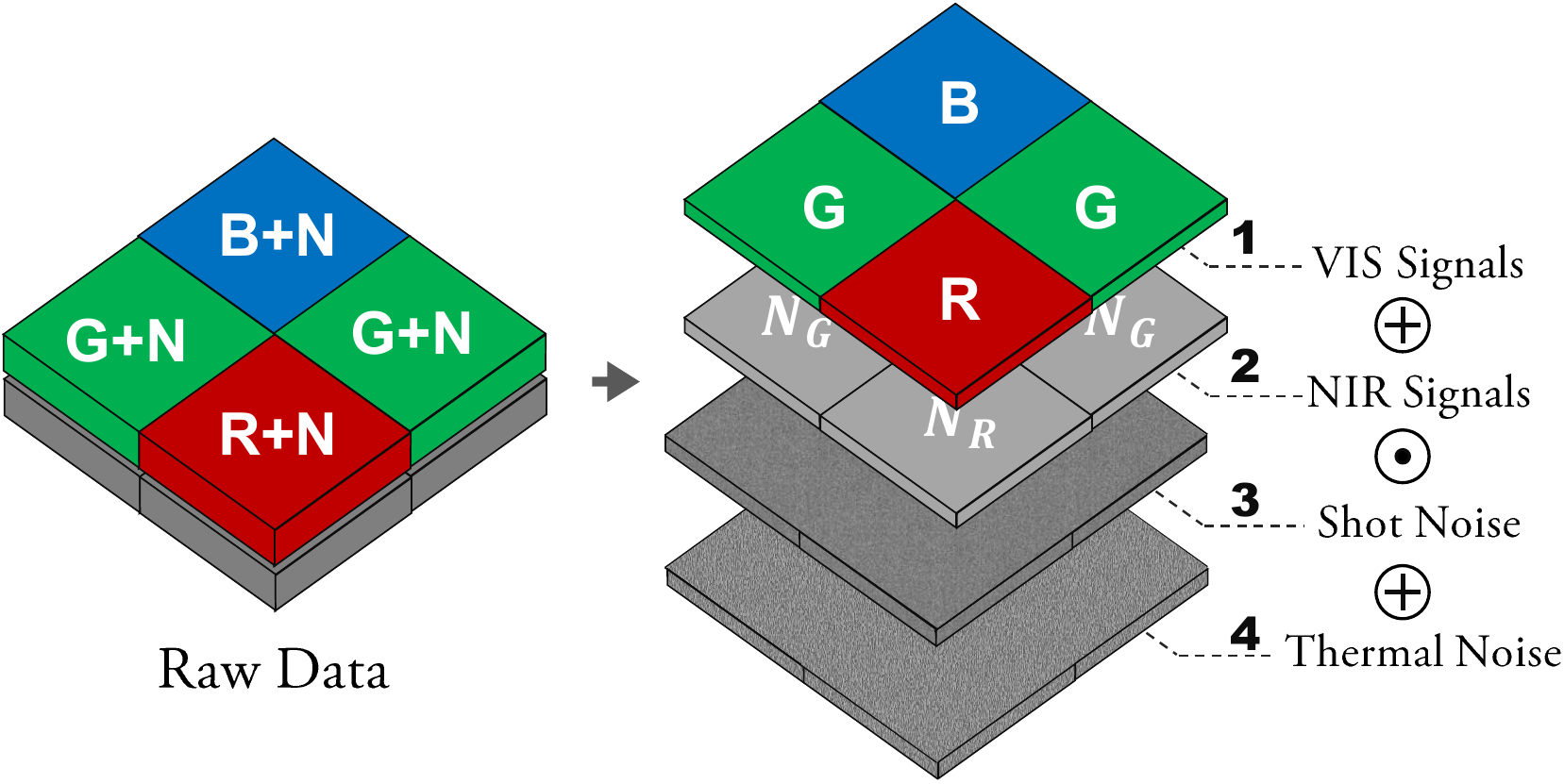}
		\end{center}
		\caption{The imaging model of our proposed solution, which consists of four main components: VIS signals, NIR signals, Shot Noise and Thermal Noise.}
		\label{fig_model}
	\end{figure}

	As described above, existing solutions are not fully satisfactory when dealing with the $24$-hour imaging task.
	In this paper, we propose a novel and complete solution for $24$-hour colorful and clear imaging, which only needs a common camera without additional hardware assistance. In this section, we will analyze the imaging mechanism of our proposed solution in detail, as shown in Figure~\ref{fig_model}.
	
	{\bf Our imaging model}.
	We consider the common camera with the hot-mirror removed so that both VIS and NIR photons can reach the sensor.
	Assuming that the automatic exposure is turned off, the raw response of one sensor element can be formulated as:
	\begin{eqnarray}
	\label{origional_model}
	S = max(0, min(1, round((g \cdot I^* + V)/\eta)),
	\end{eqnarray}
	where $g$ is the amplification factor, $V$ is the bias voltage and $\eta$ is the quantization step. The symbol $I^*$ is the released electrons and its noise-free version can be formulated as:
	\begin{eqnarray}
	\label{electron_model}
	I_0 = T A U \int_{\lambda}L(\lambda)t(\lambda)q(\lambda)d\lambda,
	\end{eqnarray}
	where $T$ is the exposure time, $A$ is the effective area, $U$ is the modulation function, $\lambda$ is the wave length, $L(\lambda)$ is the incident spectral irradiance, $t(\lambda)$ is the transmittance and $q(\lambda)$ is the photoelectric conversion function. We consider two main noise: the thermal noise and the shot noise. The real released electrons $I^*$ should be formulated as:
	\begin{eqnarray}
	I^* = I_0 + N_s(I_0) + N_t,
	\end{eqnarray}
	where $N_s(x)$ represents the shot noise according to the signal $x$ and $N_t$ represents the thermal noise. In our imaging model, $\lambda \in [300nm,950nm]$. If we use sums to approximate integrals, the VIS-NIR mixed signal can be approximated as the superposition of latent NIR and VIS signals. Thus the mixed imaging model becomes:
	\begin{eqnarray}
	\label{model}
	S_m^i  \approx  S_v^i + S_n^i + N_s(S_v^i + S_n^i) +  N_t,  i \in  \{  R,  G,  B \} ,
	\end{eqnarray}
	where $S_m$ is the captured mixed signal, $S_v$ are the latent noise-free VIS signal and $S_n$ is the latent noise-free color-filter-passed NIR signal.
	
	{\bf Analysis on noise signals}. The shot noise is dynamically affected by light intensity, while $N_t$ is usually stable and measurable. As reported in previous researches~\cite{foi2008practical,guo2018toward}, Gaussian/Poisson noise are usually adopted to approximate the thermal/shot noise.
	Assuming that the VIS signal are independent with the NIR signal, the shot noise $N_s(S_v^i + S_n^i) \approx N_s(S_v^i) + N_s(S_n^i)$.
	Following this idea, we model the noise as:
	\begin{eqnarray}
	\label{noise}
	S_v\!+\!N_s(\!S_v\!)\! \sim\! \mathcal{P}\!(\!S_v\!), \!S_n\!+\!N_s(\!S_n\!)\!\sim\!\mathcal{P}\!(\!S_n\!), \!N_t\! \sim\! \mathcal{N}\!(0, \!\sigma^2),\!
	\end{eqnarray}
	where $\mathcal{P}(\lambda)$ represents a Poisson distribution with the parameter $\lambda$, $\mathcal{N}(0, \sigma^2)$ denotes a zero-mean Gaussian distribution with variance $\sigma^2$, $S_v$ and $S_n$ are VIS signals and NIR signals respectively. By considering the simplified reflected light model, we have
	\begin{eqnarray}
	\label{lambda}
	S_v = \mathcal{F}(L(v)),~~ L(v)\approx k_v\cdot \mathcal{I}_v(t),
	\end{eqnarray}
	where $\mathcal{F}$ is the nonlinear camera response process for turning illumination into signals, $k_v$ is the reflection coefficient and $\mathcal{I}_v(t)$ represents the VIS illumination levels in time $t$. Similarly, the signal of NIR $S_n$ has the same relationship with the NIR illumination levels $\mathcal{I}_n$.
	
	By now, Eq.~\ref{noise} and Eq.~\ref{lambda} model the two major components, namely the $S_v$ and $S_v$ in Eq.~\ref{model}, which clearly rely on $\mathcal{I}_v$ and $\mathcal{I}_n$, respectively. As for $\mathcal{I}_v$, it varies significantly during a day and causes drastic change in $S_v$. It will have low signal-to-noise (SNR) ratio when $\mathcal{I}_v$ is insufficient.
	However, $\mathcal{I}_n$ is a stable large value with the presence of additional NIR illumination, and thus $S_n$ is stable and has high SNR.
	
	{\bf The final solution}. Given the above analysis, we propose to separate $S_v+N_s(S_v)$ and $S_n+N_s(S_n)$, and enhance the VIS signal with the assistance of the high-quality NIR signal. The separation is possible thanks to the stability of $S_n+N_s(S_n)$. Then the idea is that, in the daytime when both $\mathcal{I}_v$ and $\mathcal{I}_n$ are large, the noises are negligible, while in the nighttime when $\mathcal{I}_v$ becomes tiny and $\mathcal{I}_n$ is still large, the problem becomes restoring the VIS signal with information, \textit{e.g.}, scene structure, provided by the high-quality NIR signal. In this manner, the integrated solution produces high-quality images in $24$ hours. 
	To the best of our knowledge, this is the first solution of this kind.
	
	\section{Hardware Design \& Dataset Construction}
	We design a novel optical imaging system and build a new dataset for training and benchmarking.
	
	\subsection{Optical Design}
	There are two feasible ways to collect paired low/normal light images. One way~\cite{chen2018learning} is to shoot at night and using longer exposure images as reference. Another approach~\cite{Chen2018Retinex} is to imaging by adjusting the exposure time and ISO at daytime.
	The former is more consistent with our task. Therefore, we choose the first way to collect images. Specifically, at nighttime, images of each scene need to be captured in four configurations: (1) without filters, (2) with VIS pass filters, (3) with long exposure and VIS pass filters, (4) with NIR pass filters. As for the daytime, we only need to capture three images.
	
	\subsection{Hardware Configuration}
\begin{figure}[t]
		\begin{center}
			\includegraphics[scale=0.78]{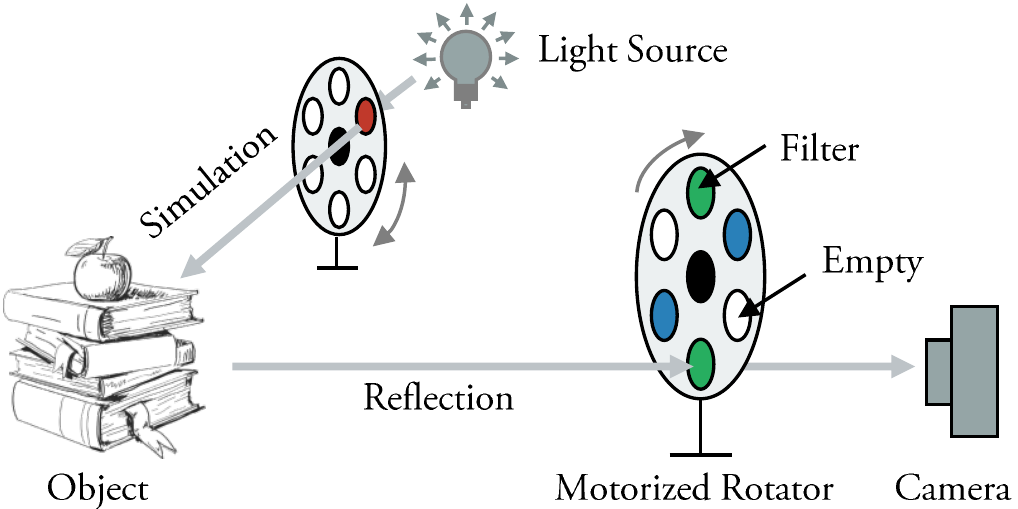}
		\end{center}
		\caption{For data collection, our designed system consists of a xenon lamp, a camera and a motorized rotator with four filters (two NIR cut-off filters and two VIS pass filters).}
		\label{fig_Schematic_Illustration}
	\end{figure}

	The schematic illustration of our designed imaging system are shown in Figure~\ref{fig_Schematic_Illustration}.
	The exposure is separately set to $0.8$s/$0.08$s for nighttime/daytime, and set to $8$s for long exposure images.
	The resolution of the used camera (Bitran CS-63C, without IR cut-off filters) is $2048\times3072$ and the pixel depth is $12$~\emph{bit}.
	We use a motorized rotator (Thorlabs FW102C) for automatic filter switching to ensure that every set of images captured by the camera is aligned. It has six holes: two placed VIS pass filters (Thorlabs FESH0700), two placed NIR pass filters (Thorlabs FELH0850), and two remained empty.
	We use a xenon lamp (Asahi Lax103) as the light source. For nighttime, we use an additional $880$~\emph{nm} bandpass filter (FWHM=$15$~\emph{nm}) to simulate the LED lighting. No filters are used for daytime.
	
	\subsection{VIS-NIR-MIX Dataset}
	We propose a novel dataset containing the well aligned VIS images, NIR images and their mixed images.
	To the best of our knowledge, this is the first real dataset of this kind, which can be considered one of our key contributions in this paper.
	Such a VIS-NIR-MIX dataset can guide the separation of the VIS and NIR signals from their mixed image, and thus enables the training of our imaging model.
	
\begin{figure}[b]
		\begin{center}
			\includegraphics[scale=0.436]{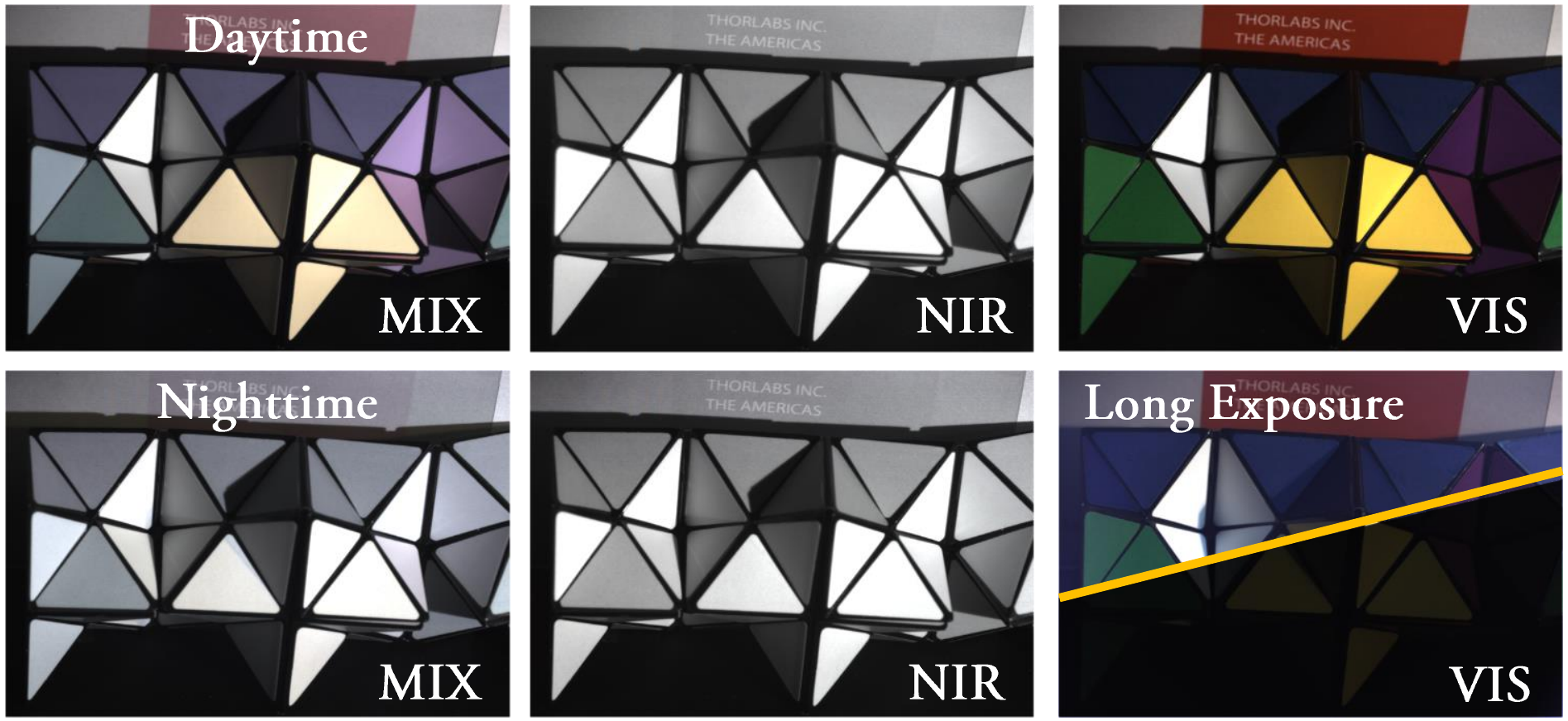}
		\end{center}
		\caption{Example images of the VNM dataset. Each scene contains $7$ images: three $3$ images and $4$ nighttime images.}
		\label{fig_dataset}
	\end{figure}

	As mentioned earlier, we collect seven images for every scene, three of which are for the daytime and four for the nighttime. Examples of the captured image set are shown in Figure~\ref{fig_dataset}.
	In addition to ensuring the physical alignment, we also manually check every sample to achieve meticulous calibration.
	By using our capture system introduced above, we totally collect $714$ images from $102$ different scenes.
	In this paper, we use $14$ images for test and the rest for training and the rest images as the training set.

	\section{Methodology}
	A specific implementation of the proposed solution is introduced in this section with all the necessary details.
	Our solution contains two major steps: separation and restoration. We design a novel network architecture consists of four sub-Nets corresponding to the two steps.
	
	\subsection{Network Architecture}
	
\begin{figure*}[t]
		\begin{center}
			\includegraphics[scale=0.20]{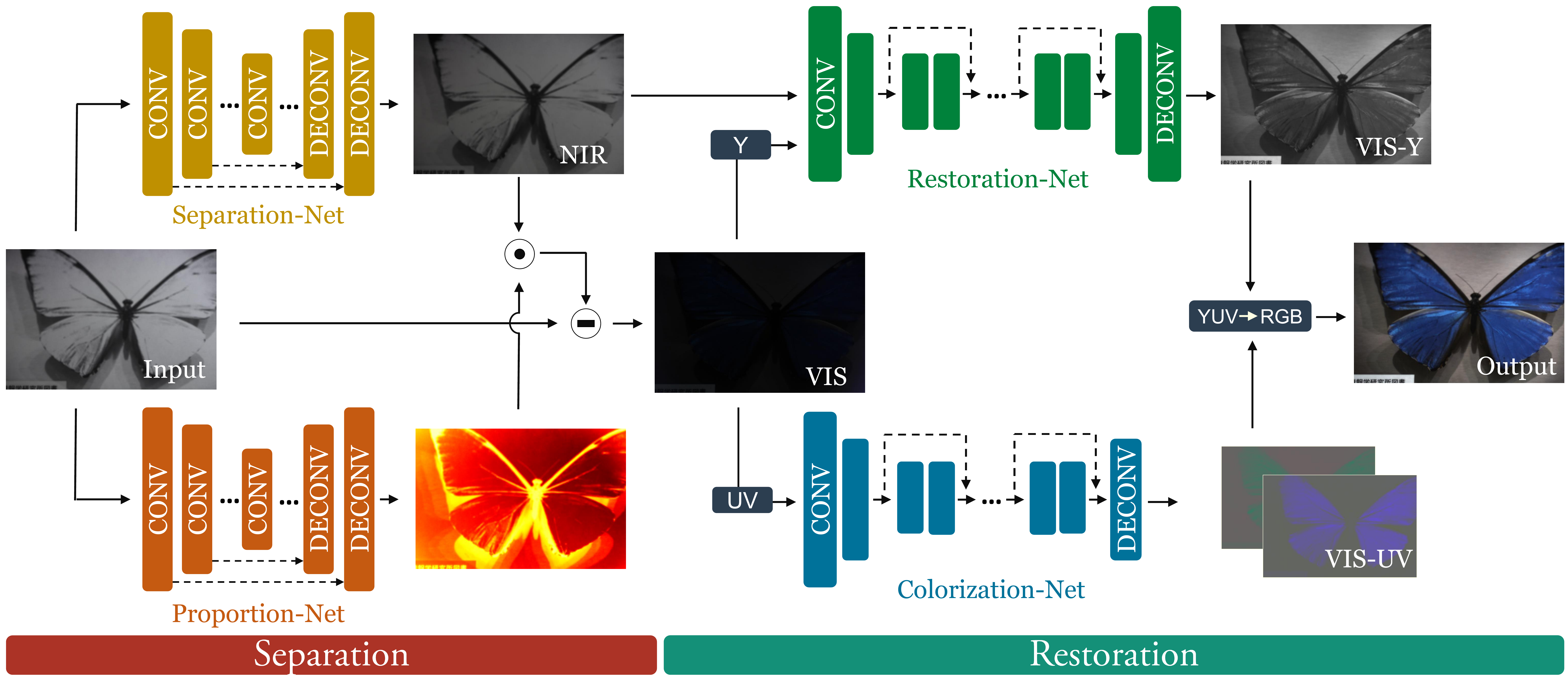}
		\end{center}
		\caption{The proposed network architecture. The Separation-Net is used to separate pure NIR signals from mixed signals. The Proportion-Net is used to estimate the deviation of the NIR signal. The Restoration-Net is used to restore the luminance component. The Colorization-Net is used to restore the chrominance information. The dashed lines represent skip connections.}
		\label{fig_network}
	\end{figure*}

	As shown in Figure~\ref{fig_network}, the proposed network consists of four sub-networks. We use U-Net~\cite{ronneberger2015u,pix2pix2017} and Res-Net as the basic element of our network since they have been proven extensively effective.
	We use Instance Normalization~\cite{ulyanov2016instance} rather than Batch-Normalization~\cite{ioffe2015batch} for Normalization.
	For the U-Net module, we use convolutional layers with stride $2$ for down-sampling and use resize-convolutional layers for up-sampling~\cite{odena2016deconvolution}. 
	As the decease of the feature map size, the number of feature maps doubles and the feature map number of the first layer is set to 64. For all layers, using Leaky-ReLU (except Proportion-Net), the kernel size is set to 3$\times$3 and the concatenating along the channel dimension instead of adding directly is used for skip connections.
	
	{\bf Separation-Net}. We use one U-Net module as the virtual NIR pass filter to generate pure NIR images. The input is the mixed image and the output is the estimated NIR image.
	
	{\bf Proportion-Net}. Due to the interval between VIS/NIR pass filters, the transmittance limitation, etc., the sum of captured NIR signals and VIS signals are not exactly equal to the mixed signal. Considering that the major components of deviation are in 700-850~\emph{nm} and are similar to NIR signals. We use one U-Net module to predict this deviation.
	Different from other sub-Nets, we use Sigmoid in the last layer and use ReLU in other layers.
	The input should be scaled to [0, 1] and the value of the output is also in [0, 1], which presents the deviation percentage of the NIR signal.
	
	{\bf Restoration-Net}. As the luminance component is similar to the NIR images, we separate the chrominance component to avoid possible interference. We directly use a ResNet-like subnet to enhance and denoise the luminance component. The inputs are the estimated NIR image and the luminance component of the estimated VIS image. The output is the restored luminance component guided by the NIR image.
	
	{\bf Colorization-Net}. Since color is much lower spatial frequency than intensity, a certain degree of compression won't result in the loss of chrominance information, but rather facilitates denoising and enhancement. Unlike other sub-Nets, the output size of Colorization-Net is only half of the input.
	The obtained color components are resized and fused with the luminance component to obtain the final result.
	
	\subsection{Loss Function}
	The total loss function we proposed is formulated as:
	\begin{align}
	\label{total_loss}
	\mathcal{L} = \alpha \cdot \mathcal{L}_{s} + \beta \cdot \mathcal{L}_{r},
	\end{align}
	where $\mathcal{L}_{s}$ and $\mathcal{L}_{r}$ represent the loss function of Separation and Restoration processes, $\alpha$ and $\beta$ are the corresponding coefficients. The details of two loss are given below.
	
	To maintain structural consistency, we use a structure-aware smoothness loss and the well-known image quality assessment algorithms SSIM along with $mae$ as the loss function. The value ranges of SSIM is $(-1,1]$ and the definition can be found in~\cite{wang2004image}. The $\mathcal{L}_{s}$ is defined as:
	\begin{align}
	\label{Sep_loss}
	\mathcal{L}_{s} \!=\! \lambda_{v} \!( \!\mathcal{L}_{ma}^{v} \!+\! 1 \!-\! \mathcal{L}_{ss}^v\!)\!+\!\lambda_{n} \!( \!\mathcal{L}_{ma}^{n} \!+\! 1 \!-\! \mathcal{L}_{ss}^n\!)\!+\!\gamma_1\! \mathcal{L}_{sm}^{n},
	\end{align}
	where $\mathcal{L}_{ma}$, $\mathcal{L}_{ss}$, $\lambda$ represents the $mae$, $ssim$ value and coefficient. The $\mathcal{L}_{sm}^{n}$ is defined as:
	\begin{align}
	\label{smooth_loss}
	\mathcal{L}_{sm}^{n} = \Vert \nabla I_{n} \circ exp(-\lambda_g\cdot \nabla I_{m}) \Vert,
	\end{align}
	where $I_{n}$ is the output of the Separation-Net, $I_{m}$ is the input image, $\nabla$ represents the gradient, $\lambda_g$ denotes the coefficient balancing the strength of structure-awareness.
	We use similar loss function for Restoration processes,
	\begin{equation}
	\begin{aligned}
	\label{Res_loss}
	\mathcal{L}_{r}=&\lambda_{v2}\mathcal{L}_{ma}^{v2}+ \lambda_{y}\mathcal{L}_{ma}^{y} +\lambda_{v2}( 1 - \mathcal{L}_{ss}^{v2})+\\
	&\gamma_2 \mathcal{L}_{sm}^{y}+ \gamma_3 \mathcal{L}_{sm}^{v2}+ \gamma_4 \mathcal{L}_{pe}^{v2},
	\end{aligned}
	\end{equation}
	where $v2$ and $y$ represent the output in RGB and YUV color spaces, $\mathcal{L}_{sm}^{y}$ and $\mathcal{L}_{sm}^{v2}$ are the smooth constrain for Luminance and VIS signals guided by NIR signals (similar to the formula~\ref{smooth_loss}), $\gamma_2,\gamma_3$ and $\gamma_4$ are the coefficient, $\mathcal{L}_{pe}^{v2}$ is the perceptual loss and the define can be find in~\cite{johnson2016perceptual}.
	In the experiments, the configuration of the parameters is: $\alpha\!=\! 1$, $\beta\!=\!1$, $\lambda_{n}\!=\!\lambda_{v}\!=\!\lambda_{v2}\!=\!\lambda_{y} \!=\!10^2$, $\gamma_1\!=\!0.1$, $\gamma_2\!=\!5$, $\gamma_3\!=\!1$, $\gamma_4\!=\!10^2$ and $\gamma_g\!=\!10$.
	
	\subsection{Implementation Details}
	Our implementation is done with Keras~\cite{chollet2015keras} and Tensorflow ~\cite{abadi2016tensorflow}. The proposed network can be quickly converged after being trained for 3000 epochs on a Titan-Xp GPU using the proposed dataset. We use random clipping, flipping and rotating for data augmentation. We set the batch-size to 10 and the size of random clipping patches to $256\times256$. The input image values are scaled to $[-1, 1]$. In the experiment, training is done using the Adam optimizer~\cite{kingma2014adam} with $\alpha = 0.0002$, $\beta_{1} = 0.9$, $\beta_{2} = 0.999$ and $\epsilon = 10^{-8}$. In the training, we reduce the learning rate to $50\%$ when the loss metric has stopped improving.
	
	\section{Experiments}
	To demonstrate the performance of our solution, we first evaluate and compare with existing methods through extensive experiments form both qualitative and quantitative aspects. Then, we evaluate the robustness in real-world scenarios under challenging conditions, including different hardware and illumination. Next, we perform the ablation experiment to evaluate the effect of different elements in our solution. Finally, we discuss the failure cases. All experiments use the model trained by using daytime and nighttime scenes simultaneously. All visualization results are white balanced.

	\subsection{Performance Comparison}
	We conduct a comprehensive experiment to prove the performance of our solution by comparing with existing representative methods. These comparison methods stand out from 20 related methods in the pre-experiment.
	Specially, for daytime scenes, we compare with Gamma correction, histogram equalization (HE) and NPE~\cite{wang2013naturalness}.
	For nighttime scenes, we compare with the extremely low-light raw image restoration method~\cite{chen2018learning}, low-light enhancement method~\cite{guo2017lime} with single image denoising method~\cite{dabov2007color} or joint denoising method~\cite{yan2013cross}, and colorization methods~\cite{iizuka2016let,zhang2016colorful}.
	Notice that, all method as well as ours need only one image as the input except joint denoising method.
	
\begin{figure}[t]
		\begin{center}
			\includegraphics[scale=0.36]{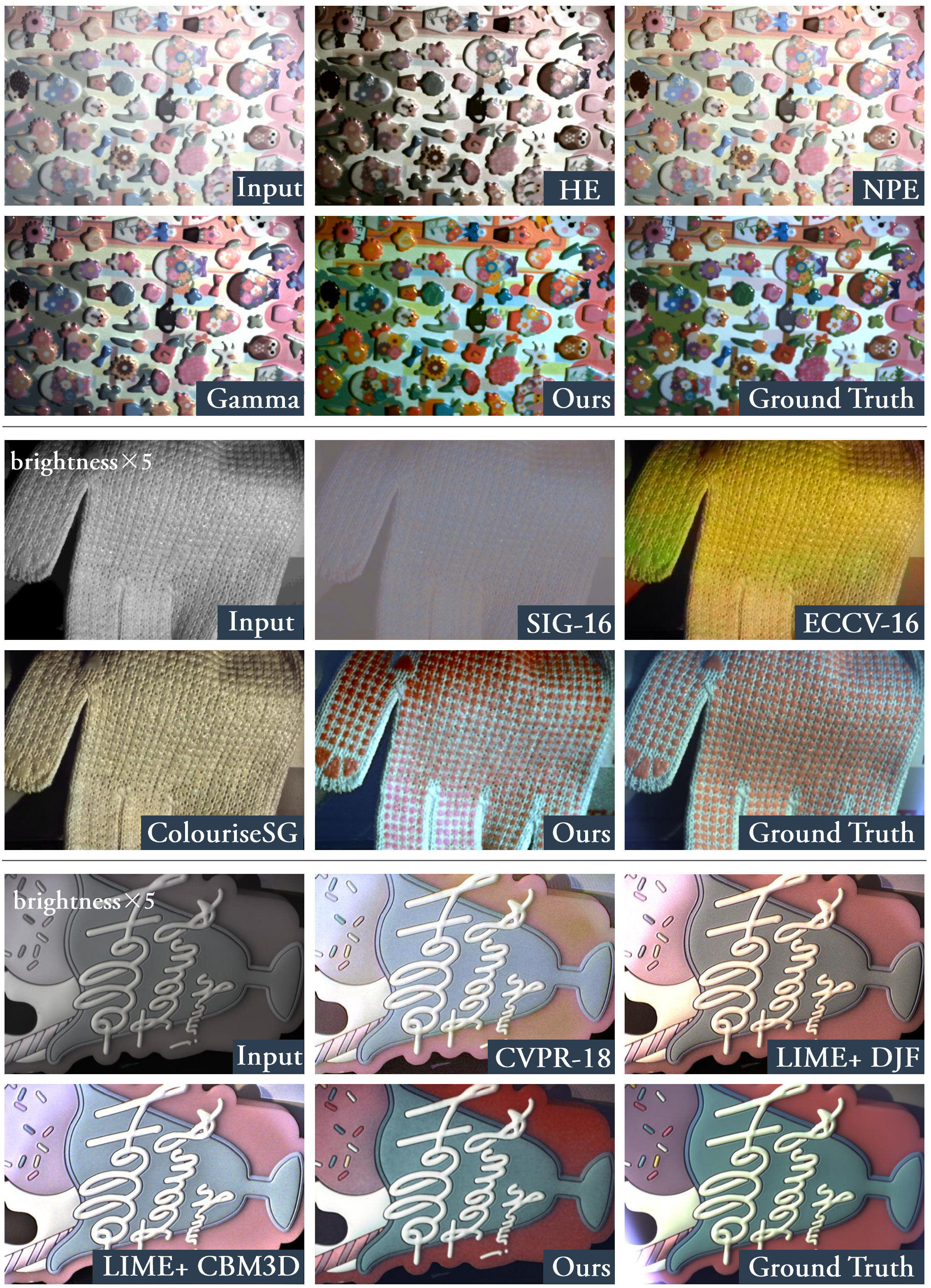}
		\end{center}
		\caption{(Top) Comparison with enhancement methods at daytime. (Middle) Comparison with colorization methods at nighttime. (Bottom) Comparison with low-light image restoration solutions at nighttime.}
		\label{fig_ex1}
	\end{figure}

	{\bf Perceptual comparison with enhancement method at daytime.} The main challenges of imaging at daytime using cameras without NIR cut-off filters are the low-contrast and low-saturation. Therefore, we compare our solution with three representative enhancement algorithms, as shown in top section of Figure~\ref{fig_ex1}. The colorful result shows the effectiveness of our solution at daytime.
	
	{\bf Perceptual comparison with the low-light enhancement and denoising methods at nighttime.} The main challenges of imaging at nighttime are the low-exposure, strong noise and other degradation. We compare with three representative solutions: low-light raw image restoration method, low-light rgb image enhancement method with single image denoising methods and NIR-guided joint restoration algorithms (need high-quality NIR images as the guidance). Figure~\ref{fig_ex1} shows that our solution can correctly restore the chrominance and our result is best in terms of color fidelity.
	
	{\bf Perceptual comparison with colorization methods at nighttime.} Directly predicting the chrominance form the grayscale NIR images seems to be a perfect solution. However, as the color is often inherently ambiguous, it is almost impossible to infer brilliance information from grayscale information in an unqualified condition, such as colorization for the Rubik's cube. We use representative colorization solutions to demonstrate the limitations of colorization methods in $24$-hour imaging. Figure~\ref{fig_ex1} shows that our solution can produce realistic images by using NIR-VIS mixed signals.
	
	{\bf Quantitative evaluation.} We use PSNR, SSIM~\cite{wang2004image} and colourfulness~\cite{hasler2003measuring} as the image quality metrics. Table~\ref{tab_enhancement} shows that our solution outperforms all comparison methods in all quantitative metrics whether at daytime or nighttime. Particularly, we use only one model for imaging in both daytime and nighttime, which demonstrates the effectiveness of our solution.
	
	\begin{table}[htbp]
		\centering
		\small
		\begin{tabular}{l|cccl}
			\hline\hline
			~ & PSNR & SSIM &Color & Runtime \\ \hline\hline
			Input-daytime & 17.66 &  0.714 & 38.43 & -\\
			HE & 16.38  &  0.691 & 39.16 & {\bf 1.53}\\
			NPE & 16.59  &  0.703 & 38.68 & 188.68\\
			Ours & {\bf 22.25}  &  {\bf 0.871} & {\bf 73.24} & 4.03 ~~(GPU)\\ \hline\hline
			Input-nighttime & 14.73 &  0.632 & 67.58 & -\\
			CVPR18 & 15.32  & 0.678 & 39.00 & 3.50 ~~(GPU)\\
			LIME+CBM3D & 15.52  &  0.756 & 55.26 & 157.66\\
			LIME+DJF & 15.64  &  0.739 & 59.97 & 91.82\\
			SIG16 & 11.16  &  0.056 & 14.65 & 58.25\\
			ECCV16 & 13.25  &  0.575 & 61.35 & 16.56 (GPU)\\
			Ours & {\bf 17.17}  &   {\bf 0.835} & {\bf 89.74} & 4.03 ~~(GPU)\\
			\hline\hline
		\end{tabular}
		\caption{Quantitative evaluation comparative result. The size of images for testing runtime is $3072\times2048$. The top (bottom) half is the average metric results of daytime (night).}
		\label{tab_enhancement}
	\end{table}

	\subsection{Robustness Verification}
	We validate the robustness of our solution by testing three challenging cases: imaging under different illumination conditions, imaging for outdoor scenes under real sunlight, and imaging using different cameras.
	
\begin{figure}[htbp]
		\begin{center}
			\includegraphics[scale=0.38]{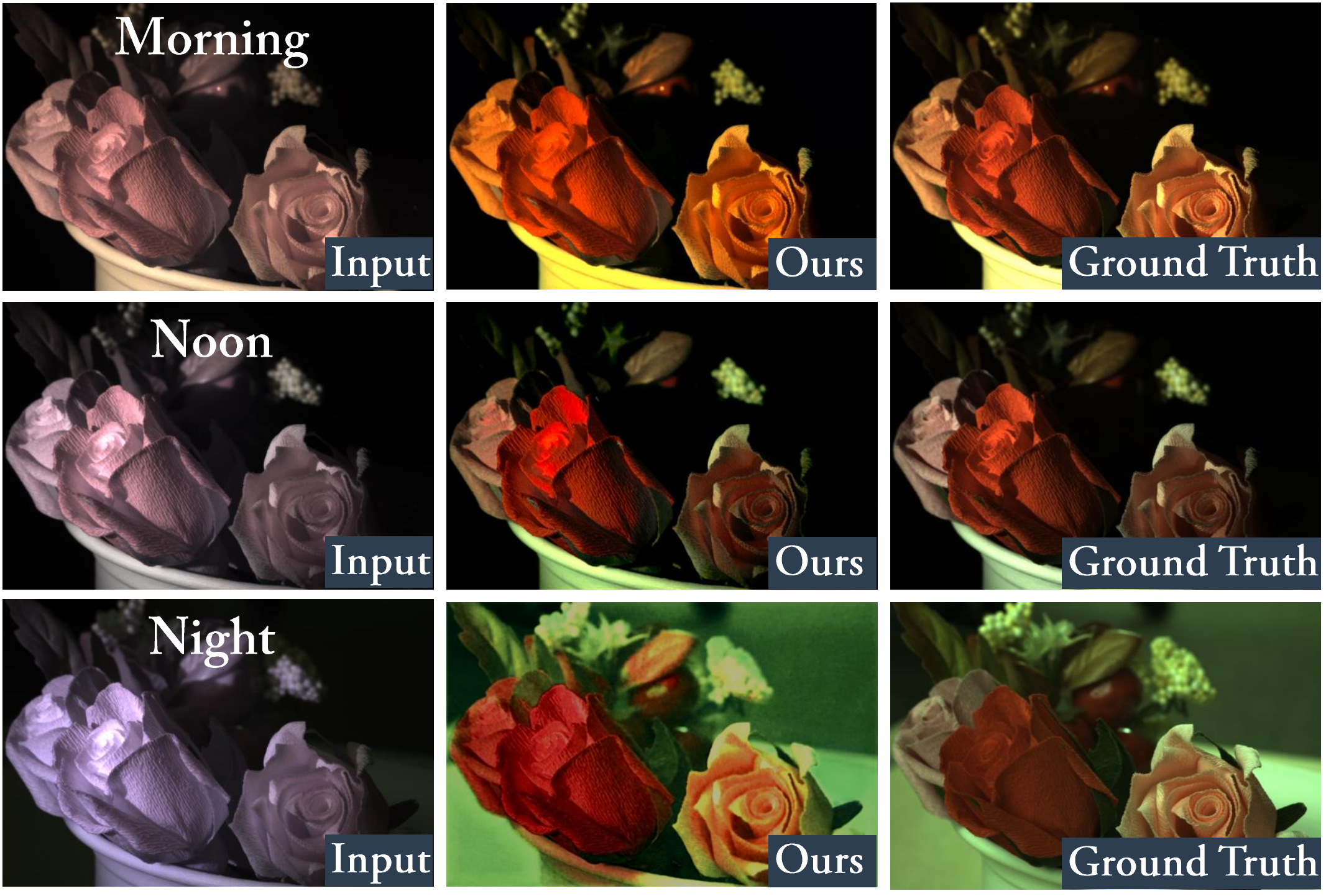}
		\end{center}
		\caption{Our solution restores images correctly at different times under different illumination conditions.}
		\label{fig_ex2_illumination}
	\end{figure}

	{\bf Different illumination conditions.} To demonstrate the robustness of our solution in restoring images in different illumination conditions, we provide results of real scenes at different times to verify the effectiveness of our solution. Figure~\ref{fig_ex2_illumination} shows that our solution can correctly restore images under different illumination conditions.
	
\begin{figure}[b]
		\begin{center}
			\includegraphics[scale=0.38]{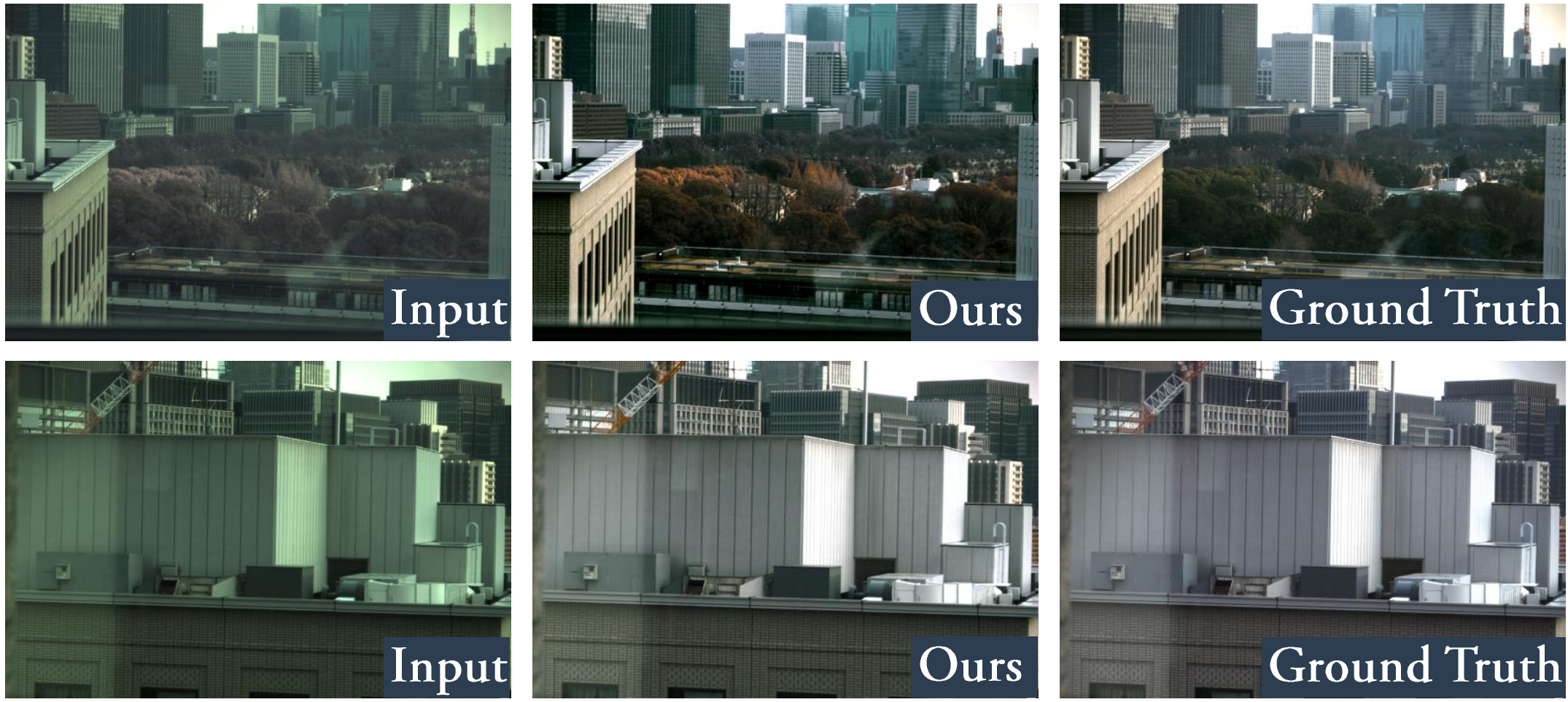}
		\end{center}
		\caption{Our solution correctly enhances outdoor images under real sunlight.}
		\label{fig_ex2_sunlight}
	\end{figure}

	{\bf Outdoor scenes under real sunlight.} Although our image dataset only contains indoor scenes under artificial illumination, our models can enhance outdoor scenes under real sunlight correctly as shown in Figure~\ref{fig_ex2_sunlight}. In addition, thanks to the mixed training using daytime and nighttime scenes simultaneously, our enhancement results are better than imaging using NIR cut-off filters in terms of contrast.
	
\begin{figure}[t]
		\begin{center}
			\includegraphics[scale=0.38]{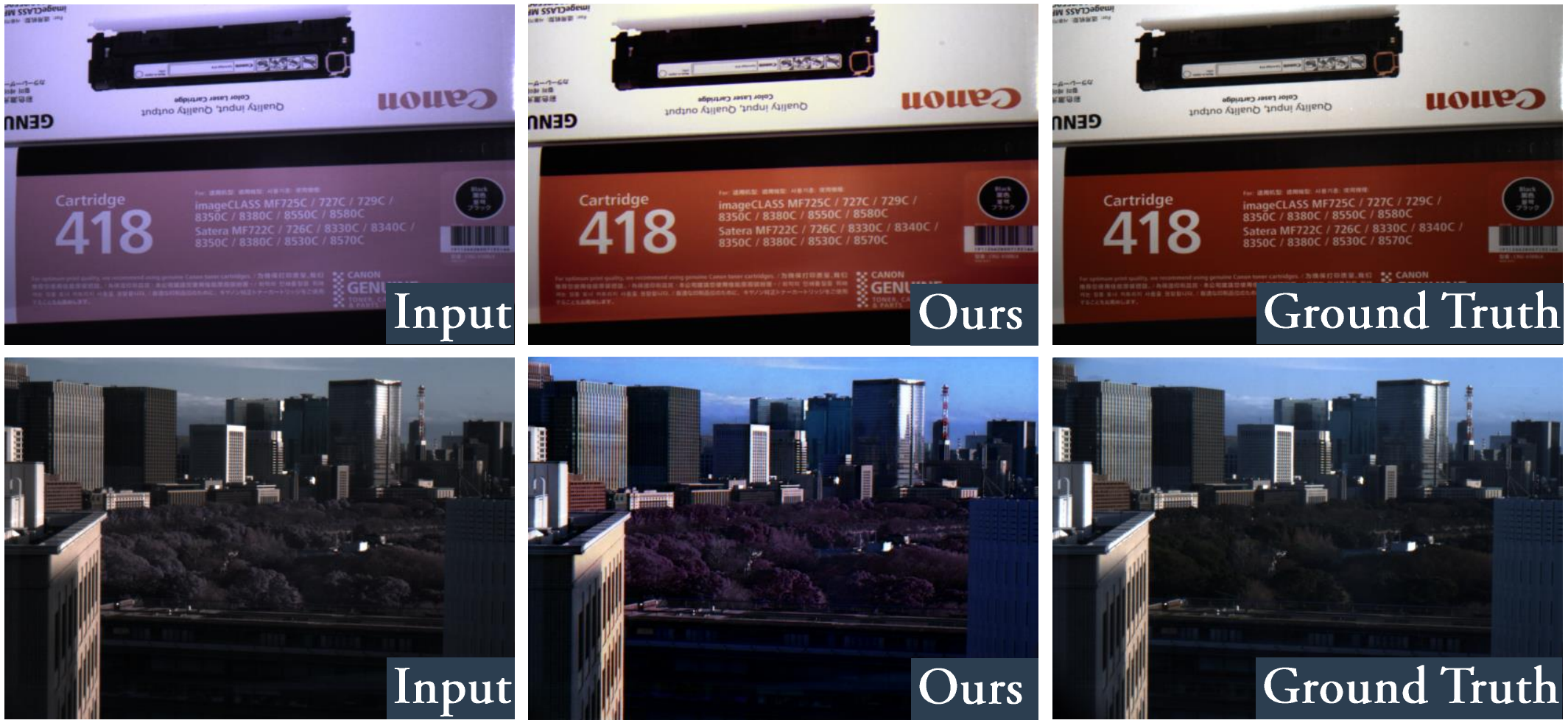}
		\end{center}
		\caption{Our solution correctly enhances indoor and outdoor images captured by using low-end cameras at daytime.}
		\label{fig_ex2_difcamera}
	\end{figure}

	{\bf Different cameras.} Robustness across extremely different hardware is a challenging task, as imaging colors are influenced by the camera spectral sensitivity~\cite{Gao:17}. Fortunately, as for different sensors, spectral sensitivity dataset~\cite{Jiang_whatis} has shown that various Silicon cameras have similar sensitivities curves, which demonstrates that our model can be directly applied to low-end cameras to some extent. To certificate this, we use a low-end camera (FLIR GS3-U3-15S5C-C) for testing. Its resolution is $1032\times1384$ and the pixel depth is $8$~\emph{bit}.
	The result using the pre-trained model without any fine-tuning is shown in Figure~\ref{fig_ex2_difcamera}. This demonstrates the robustness of our model across hardware.
	
	\subsection{Ablation Experiment}
	In this section, we quantitatively evaluate the effectiveness of different components in our solution based on our VNM dataset. Table~\ref{tab_control} reports the performance of the presented changes in terms of PSNR and SSIM~\cite{wang2004image}.
	
	{\bf Loss functions}. We evaluate the impact of loss function components as shown in Table~\ref{tab_control} (row $2-4$).
	It shows that the result quality is improving by containing more loss components, which demonstrates the effectiveness of our proposed loss function from the side.
	
	{\bf Network structures}. As shown in Table~\ref{tab_control} (row $5-8$), we evaluate the effectiveness of different network modules by replacing using a U-Net. Similar to the loss function, the results show that more components of our network will result in better performance. Besides, modifying the Colorization-Net's output size will slightly decrease the performance.
	
	{\bf Color space}. Table~\ref{tab_control} (row $9-10$) shows the impact of the color space. Directly restoring in HSV (use network like condition $8$) and RGB (use U-Net replace the restoration module) color space will seriously reduce the result quality. 
	
	\begin{table}[htbp]
		\begin{tabular}{l|ccc}
			\hline \hline
			Condition & PSNR &SSIM  \\ \hline\hline 
			1. Our full model  &  {\bf 19.71} & {\bf 0.853} \\\hline\hline
			2. w/i $\mathcal{L}_{ma}$, w/o $\mathcal{L}_{ss}$, w/o $\mathcal{L}_{sm}$, w/o $\mathcal{L}_{pe}$ & 18.71 & 0.822 \\
			3. w/i $\mathcal{L}_{ma}$, w/i $\mathcal{L}_{ss}$, w/o $\mathcal{L}_{sm}$, w/o $\mathcal{L}_{pe}$ & 19.12 & 0.846 \\
			4. w/i $\mathcal{L}_{ma}$, w/i $\mathcal{L}_{ss}$, w/i $\mathcal{L}_{sm}$, w/o $\mathcal{L}_{pe}$ & 19.54 & 0.839 \\
			\hline\hline
			5. w/o Separation, w/o Restoration & 16.74  &  0.801\\
			6. w/i Separation, w/o Restoration & 18.88 & 0.826\\
			7. w/o Separation, w/i Restoration & 17.06 & 0.811\\
			8. Colorization-Net's output size $\times 2$ & 19.10 & 0.836\\
			\hline\hline
			9. YUV $\rightarrow$ HSV color space  & 18.30 & 0.806 \\
			10. YUV $\rightarrow$ RGB color space &  17.58 & 0.782 \\
			\hline\hline
		\end{tabular} \\
		\caption{Ablation experiment results. This table report mean PSNR/SSIM in each condition on the test dataset. In this table, "w/i" means with and "w/o" means without.}
		\label{tab_control}
	\end{table}

	\section{Conclusion \& Discussion}
	In this paper, we propose a novel solution for 24-hour high-quality imaging. We propose a new imaging model, design a multi-channel imaging processing system and build a new dataset called VIS-NIR-MIX Dataset (VNM). Using the new dataset, we design a proof-of-concept prototype using a fully-convolution network. Extensive experiments demonstrate that our solution is effective and practical.
	
	{\bf Benefits.} Our solution can be used for 24-hour high-quality imaging, and it avoids the use of additional hardware like specific CFAs, two cameras, and so on. 
	
	{\bf Limitations.} Our solution is not specifically optimized for multi-camera consistency. As distributions of signals captured using diverse cameras are different, maintaining robust across different cameras is challenging, especially in extremely low-light conditions. Collecting more diverse data is a promising solution and we leave this for the future work.
	
	{\bf Future work.} Future works may focus on improving the multi-camera consistency, handling dynamic scenarios, and optimization for specific applications like surveillance. In addition to improving the model performance, laborious data preparation is also need to be improved. How to improve the efficiency of data preparation or produce realistic synthetic data for enlarging the dataset are another aspect.
	
	\bibliography{AAAI-LvF.1088}
	\bibliographystyle{aaai}
	
\end{document}